\documentclass[11pt]{article}
\usepackage{amsmath,amssymb}
\usepackage{mathrsfs}
\usepackage{graphicx,psfrag,epsf}
\usepackage{enumerate}
\usepackage{natbib}
\usepackage[ruled]{algorithm2e}
\usepackage{tabularx}    
\usepackage{setspace}  
\usepackage{float}    
\usepackage[toc,page]{appendix}

\newcommand{\blind}{0}

\def\spacingset#1{\renewcommand{\baselinestretch}%
{#1}\small\normalsize} \spacingset{1}

\begin{document}


\if0\blind
{
  \title{Two-Stage Approach for the Inference of the Source of High-Dimensional and Complex Chemical Data in Forensic Science}
  \author{Madeline AUSDEMORE, Cedric NEUMANN, Christopher SAUNDERS, \\ Douglas ARMSTRONG and Cyril MUEHLETHALER \thanks{Madeline Ausdemore (\textit{E-mail:madeline.ausdemore@sdstate.edu}) is a Doctoral Student in Computational Science and Statistics, Cedric Neumann (\textit{E-mail: cedric.neumann@sdstate.edu}) is an Associate Professor of Statistics, Christopher Saunders (\textit{E-mail: christopher.saunders@sdstate.edu}) is an Associate Professor of Statistics, and Douglas Armstrong (\textit{E-mail: douglas.armstrong@sdstate.edu}) is a Ph.D. graduate in Computational Science and Statistics, Department of Mathematics and Statistics, South Dakota State University, Brookings, SD 57007-3511. Cyril Muehlethaler (\textit{E-mail: cyril.muehlethaler@uqtr.ca}) is an Assistant Professor of Criminalistics, D\'epartement de Chimie, Biochemie et Physique, Universit\'e du Qu\'ebec \`a Trois Rivi\`eres, Trois Rivi\`eres, QC G9A 5H7, Canada. This project was supported in part by Award No. 2014-IJ-CX-K088 awarded by the National Institute of Justice, Office of Justice Programs, U.S. Department of Justice. The opinions, findings, and conclusions or recommendations expressed in this paper are those of the authors and do not necessarily reflect those of the U.S. Department of Justice.}}
  \maketitle
} \fi
\if1\blind
{
  \bigskip
  \bigskip
  \bigskip
  \begin{center}
    {\textbf{Two-Stage Approach for the Inference of the Source of High-Dimensional and Complex Chemical Data in Forensic Science}}
\end{center}
  \medskip
} \fi

\bigskip
\begin{abstract}
Forensic scientists are often criticised for the lack of quantitative support for the conclusions of their examinations. While scholars advocate for the use of a Bayes factor to quantify the weight of forensic evidence, it is often impossible to assign the necessary probability measures to perform likelihood-based inference for high-dimensional and complex data. To address this issue, we revisit a two-stage inference framework and leverage the properties of kernel functions to offer a method that allows for statistically supporting the inference of the identity of source of sets of trace and control objects by way of a single test. Our method is generic in that it can be easily tailored to any type of data encountered in forensic science or pattern recognition, and our method does not depend on the dimension or the type of the considered data. The application of our method to paint evidence shows that this type of evidence carries substantial probative value. Finally, our approach can easily be extended to other evidence types such as glass, fibres and dust.
\end{abstract}

\noindent%
\vfill

\newpage
\spacingset{1.75} 

\section{Introduction}
\label{sec:intro}
Given $M$ trace objects assumed to originate from a single source, and $N$ control objects from a known source, we want to infer if all $N+M$ objects originate from the same source. Formally, we want to test if:
\begin{itemize}
\item[] $H_1$ - the set of $M$ trace objects and the set of $N$ control objects are two simple random samples from the source of the $N$ control objects;
\item[] $H_2$ - the set of $M$ trace objects is a simple random sample from another source in a population of potential sources. 
\end{itemize}
In forensic science, differentiating between these two propositions cannot be reduced down to a simple classification or model selection problem that can be directly addressed by machine learning or other similar techniques: 
\begin{enumerate}[1.]
\item The required approach needs to address the two competitive hypotheses, $H_1$ and $H_2$, presented above, from the comparison of two groups of objects (and by learning the characteristics of the source of the control objects from a training sample and performing multiple dependent tests to determine if each individual trace object can be associated with that source as proposed by \cite{ASTM:2926-13} and \cite{Park:2018}).
\item The required approach needs to account for the potentially limited number of samples available to the forensic scientist (e.g., 3-10 observations); hence techniques requiring extensive training are not an option.
\item Material of forensic interest, such as paint, is often subject to alteration due to their exposure to environmental conditions (e.g., sun, heat); furthermore, it is not reasonable to expect to exhaustively survey the population of paint (or glass, or fibres). This implies that classifiers relying on background training data would need to be retrained every time a new source is considered in casework.
\item As explained below, the inference process must account for the many sources in the population that are potentially indistinguishable from the source of the $N$ control objects.
\end{enumerate}

Legal and scientific scholars advocate for the use of a Bayes factor to quantify the support of the observations made on the trace and control objects in favour of one of these two propositions (see \cite{RSS1} for a comprehensive discussion). Unfortunately, it is often impossible to assign the necessary probability measures to perform likelihood-based inference for the high-dimensional and complex data commonly encountered in forensic science. For example, the random vectors associated with the chemical spectra characterising glass, paint, fibre or dust evidence may have thousands of dimensions and may include different types of data (e.g., discrete, continuous, compositional). Without these probability measures, assigning Bayes factors, or performing any other likelihood-based inference, is not possible. Consequently, forensic scientists reporting these types of evidence are left without means to support their assessment of the probative value of the evidence. 


In this paper, we revisit the \textit{two-stage inference framework} formally introduced by Parker (\citeyear{Parker:1966,Parker:1967}) and \cite{Parker:1968} by leveraging the properties of kernel functions \citep{Schoelkopf:2001} and the results presented by \cite{Armstrong:2017}. The proposed inference framework relies on a kernel function and, therefore, is particularly suited for high-dimensional, complex and heterogenous data. The framework is generic and can easily be tailored to any type of data by modifying the kernel function. Our solution involves algorithms that allow for handling the uncertainty on the model's parameters, and permit rapid and efficient sampling from the posterior distributions of these parameters. Furthermore, it relies on a single main assumption, which can be satisfied through the design of the kernel function. 

Despite some well-known shortcomings of the two-stage approach, discussed later in this paper, we believe that the proposed approach can provide a helpful and rigorous statistical framework to support the inference of the identity of source of trace and control objects described by high-dimensional heterogenous random vectors, such as chemical spectra. We used the proposed approach to study the probative value of traces of paint (characterised by Fourier-Transform Infrared spectroscopy (FTIR)) that may be observed in connection with crimes (e.g., paint present on tools used to force open doors or windows).
\section{Overview of Parker's two-stage approach}
\label{overview}
The general framework of the two-stage approach was first briefly mentioned by \cite{Kirk:1953} and \cite{Kingston:1965}, and was formally described by Parker (\citeyear{Parker:1966,Parker:1967}) and \cite{Parker:1968}. Parker breaks down the forensic inference process into two stages, which he describes as the \textit{similarity stage}, and the \textit{discrimination stage}. In the similarity stage, the goal is to compare the characteristics of the trace and control objects and determine whether they are \textit{distinguishable}. As the difference between the sets of characteristics increases, the hypothesis that the trace and control objects originate from the same source is weakened to the point that it can be rejected. However, establishing that the two sets of characteristics are indistinguishable is not sufficient in itself to conclude to the identity of the sources of the two sets of objects. Intuitively, the value of finding that the set of characteristics describing the trace objects is indistinguishable from that of the control objects is a function of the number of sources whose characteristics would also be deemed indistinguishable from the trace objects using the same analytical technique: the lack of distinguishability between trace and control objects is more valuable in cases where very few sources in a population of potential sources share the same characteristics as the trace objects. 
Thus, the goal of the discrimination stage is to determine the rarity of the characteristics observed in the first stage in the population of potential sources if the two sets of objects are found to be indistinguishable from one another. The level of rarity of the trace characteristics in a population of potential sources is often called a \textit{match probability} or \textit{probability of coincidence}. 

While occasional early uses of Bayesian inference in the judicial system have been reported before the 1960s \citep{TaroniChampod:1999}, the two-stage approach appears to be an initial attempt to formally frame the problem of the inference of the identity of source in forensic science in a logical manner, and to propose a statistically rigorous method to support this inference process. Today, the two-stage approach naturally arises as a proxy for the Bayes factor in situations where the measurements made on the trace and control objects are discrete and can easily be compared, such as in single-source forensic DNA profiling: the DNA profile of a trace is compared with that of a known individual, and, if found similar, the match probability of that profile in a population of potential donors is determined \citep{Butler:2015}. 

The two-stage approach was refined in the context of glass evidence in a series of papers starting in 1977 \citep{Evett:1977}. Today, ad-hoc implementations of the two-stage approach can notably be found in relation to glass, paint and fibre evidence (see for example \cite{Curran:1997a}; \cite{Champod:1997}; \cite{Aitken:2004}; \cite{Massonnet:2014}; \cite{Muehlethaler:2014}). In most cases, the decision to reject the hypothesis that trace and control objects are indistinguishable during the first stage is based on the \textit{training} and \textit{experience} of the forensic analyst performing the examination, and the second stage is not considered \citep{Kaye:2017}. When it is considered, the determination of the match probability relies, in the best situation, on frequency estimates obtained by determining the size of an ill-defined set of objects that are considered to have the ``same characteristics'' as those of the trace \citep{Kaye:2017}. Outside of trivial situations with discrete data (e.g., blood typing) or low dimensional continuous data (e.g., refractive index of glass), we have not found a rigorous implementation of the two-stage approach that is capable of handling high-dimensional and complex forms of evidence, such as chemical spectra or impression and pattern evidence, and we have to agree with the arguments brought forward by \cite{Kaye:2017}.

Below, we propose a formal statistical method to test the hypothesis that two high-dimensional and complex sets of observations are indistinguishable (Parker's similarity stage). We extend the work published by \cite{Armstrong:2017} to develop a generic $\alpha$-level test for comparing sets of high-dimensional, heterogenous random vectors, in which we account for the uncertainty on the model's parameters, and we propose a computationally efficient algorithm that enables to increase the number of objects considered and to improve the reliability of the test. Because our test relies on kernel functions that can be tailored to any type of data, the same test can be used in multiple situations, irrespective of the type of evidence considered. Finally, our method's main assumption can be satisfied through the design of the kernel function. 

Our method improves upon existing pattern recognition methods that could be considered for addressing this type of problem, such as Support Vector Machines, Artificial Neural Networks, or Random Forests: our method does not require a training set; it allows for comparing sets of objects to each other in a single test (as opposed to comparing individual objects in multiple dependent tests); it permits likelihood-based inference; and it enables formal statistical hypothesis testing in high-dimensional, complex and heterogenous vector spaces. 

In this paper, we apply the proposed statistical test to Fourier-Transform Infrared (FTIR) spectra of paint fragments and we propose a strategy to assess the type-I and type-II errors of the test. We also extend the method to the second stage (Parker's discrimination stage) and discuss how to assign match probabilities to sets of spectra. Finally, we discuss the benefits and limitations of the two-stage approach in the context of making inference on the source of high-dimensional complex forms of forensic evidence. 
\section{First stage: testing \textit{indistinguishability}}
\label{similarity}
In the first stage of our approach, we wish to test whether a set of $M$ trace objects is indistinguishable from a set of $N$ control objects. We use an $\alpha$-level test to address $H_1$ and $H_2$. Given the nature of the test, we can only reach one of two conclusions: 
\begin{enumerate}[1.]
\item The characteristics of the sets of trace and control objects are considered to be sufficiently different. Thus, the decision is that the objects cannot originate from the same source and $H_2$ is accepted. This decision is associated with a $\alpha$-rate of erroneously rejecting the hypothesis of common source;
\item The characteristics of the sets of trace and control objects are within some level of tolerance of each other. Thus, we do not have enough evidence to reject the possibility that the sets of trace and control objects originate from the same source, and so we fail to reject $H_1$ at the chosen $\alpha$-level. 
\end{enumerate}
We want to reiterate that, in the forensic context, the latter conclusion does not directly imply that the trace and control objects originate from the same source: it merely implies that the sources of the trace and control objects are indistinguishable from each other, based on the considered characteristics and the chosen $\alpha$-level. As mentioned above, the value of finding that these sources are indistinguishable can be assessed only in light of the number of sources that would also be found to be indistinguishable from the trace source. Assessing the rarity of the trace's characteristics is the purpose of the second of the two stages, and is discussed later in this paper. 

To statistically test $H_1$ and $H_2$ in the presence of high-dimensional, heterogenous and complex data, we extend the results presented by \cite{Armstrong:2017} (and summarised in Appendix A) to develop a statistical test using vectors of scores resulting from the cross-comparisons of the trace and control objects. 
\subsection{$\alpha$-level test for vectors of scores}
\label{alpha_test}
Given two vectors of measurements $\mathbf{x}_i$ and $\mathbf{x}_j$ representing the observations made on two objects, $i,j$, a kernel function, $\kappa$, is used to measure their level of similarity and report it as a score, $s_{i,j}=\kappa(\mathbf{x}_i,\mathbf{x}_j)$. We note that the kernel function at the core of the proposed model, $\kappa$, can be designed to accommodate virtually any type of data, and should satisfy only two requirements: it must be a symmetric function, that is $\kappa(\mathbf{x}_i,\mathbf{x}_j)=\kappa(\mathbf{x}_j,\mathbf{x}_i)$, and it must ensure that the distribution of the vector of scores $\mathbf{s}=\{s_{ij}\}$ is normally distributed to satisfy the assumption made on the score model by \cite{Armstrong:2017}. This assumption is reasonable for high-dimensional objects and can be satisfied through careful design of the kernel function \citep{ArmstrongPhD}.

Given $M$ trace objects and $N$ control objects, we define the vector of scores $\mathbf{s}_{m+n}=\binom{\mathbf{s}_m}{\mathbf{s}_n}$, where $\mathbf{s}_n$ represents the $n=\binom{N}{2}$ scores calculated between all pairs of control objects, and $\mathbf{s}_m$ represents the $m=\binom{N+M}{2}-\binom{N}{2}$ scores calculated between all pairs of objects involving at least one of the trace objects. 

Since all control objects are known to originate from a single source, we use the results in \cite{Armstrong:2017} to assume $\mathbf{s}_{n} \sim~MVN(\theta\mathbf{1}_{n},\mathbf{\Sigma}_{n \times n})$ with $\mathbf{\Sigma}_{n\times n}=\mathbf{PP}^t\sigma^2_a+\mathbf{I}_n\sigma^2_e$, and parameter $\mathbf{\Psi} =~\{\theta, \sigma^2_a,\sigma^2_e\}$, where $\theta$ is the expected value of the score between any two objects from the same considered source, $\sigma^2_a$ and $\sigma^2_e$ are the variances of the two random effects, and $\mathbf{P}$ is an $n\times N$ design matrix where each row represents an $i,j$ combination of objects and consists of ones in the $i^{th}$ and $j^{th}$ positions and zeros elsewhere.

Furthermore, under $H_1$, all trace and control objects are assumed to originate from the same source, therefore
\begin{equation}
	\begin{aligned}
		\binom{\mathbf{s}_m}{\mathbf{s}_n}|H_1 & \sim  MVN(\theta\mathbf{1}_{(m+n)},\mathbf{\Sigma}_{(m+n) \times (m+n)}) \\
		& =  MVN\Bigg(\theta \mathbf{1}_{(m+n)}, \begin{bmatrix} \mathbf{\Sigma}_{m\times m} & \mathbf{\Sigma}_{m\times n} \\ \mathbf{\Sigma}_{n \times n} & \mathbf{\Sigma}_{n \times m} \\ \end{bmatrix}\Bigg)  \\[3mm]
		& =  MVN(\theta \mathbf{1}_{(m+n)}, \mathbf{QQ}^t\sigma^2_a+\mathbf{I}_n\sigma^2_e)
		\label{joint.dist.H0}
	\end{aligned}
\end{equation}
where $\mathbf{Q}$ is a design matrix of the same construction as $\mathbf{P}$, but with dimensions corresponding to the vector $\binom{\mathbf{s}_m}{\mathbf{s}_n}$.  Under $H_1$, this distribution has the same parameter, $\mathbf{\Psi} =\{\theta, \sigma^2_a,\sigma^2_e\}$, as the distribution of $\mathbf{s}_{n}$, since the only differences between the distributions are the length of the mean vectors and the dimensions of the design matrices $\mathbf{P}$ and $\mathbf{Q}$.

We begin designing the test statistic of our $\alpha$-level test by defining the \textit{conditional likelihood} of the vector of scores involving at least one trace object, given the vector of scores involving only control objects, $\mathscr{L}(\mathbf{s}_m |\mathbf{s}_n, \mathbf{\Psi})$. We then define our test statistic as the function
\begin{eqnarray}
\label{Tfun}
	T(\mathbf{s}_m, \mathbf{s}_n, \mathbf{\Psi}) \ = \ \textbf{Pr}(\mathscr{L}(\mathbf{s}_m |\mathbf{s}_n, \mathbf{\Psi}) \geq \mathscr{L}(\mathbf{s}_{m}^* |\mathbf{s}_n, \mathbf{\Psi}))	,
	\label{define.pval}
\end{eqnarray}
where $\mathbf{s}_m^*$ is a random vector of scores calculated between pairs of objects involving at least one trace object  when the trace objects truly originate from the same source as the control objects. The distribution of $\mathbf{s}_m^*|\mathbf{s}_n, \mathbf{\Psi}$, obtained using the structure of the covariance matrix defined in (\ref{joint.dist.H0}), is 
\begin{eqnarray}
	\hspace{3mm}  \mathbf{s}_{m}^*|\mathbf{s}_{n},\mathbf{\Psi} \sim MVN(\theta \mathbf{1}_m + \mathbf{\Sigma}_{m \times n}\mathbf{\Sigma}_{n\times n}^{-1}(\mathbf{s}_n - \theta\mathbf{1}_n), \mathbf{\Sigma}_{m\times m}-\mathbf{\Sigma}_{m \times n}\mathbf{\Sigma}_{n \times n}^{-1}\mathbf{\Sigma}_{n\times m}).
	\label{conditional.dist.sm}
\end{eqnarray}
Using this test statistic, we decide to reject $H_1$ at a specific $\alpha$-level if
\begin{eqnarray}
\label{pvalue}
	T(\mathbf{s}_m, \mathbf{s}_n, \mathbf{\Psi}) \ \leq \ c(\alpha),
\end{eqnarray}
where $c(\alpha)$ is a constant chosen to satisfy  
\begin{eqnarray}
\label{Prpvalue}
	\textbf{Pr}\left(T(\mathbf{s}_m, \mathbf{s}_n, \mathbf{\Psi}) \leq c(\alpha)\right) \ \leq \ \alpha.
\end{eqnarray}
For a well-behaved test,  $c(\alpha) = \alpha$ by construction of $T(\mathbf{s}_m, \mathbf{s}_n, \mathbf{\Psi})$. In practice,  there is uncertainty about $\mathbf{\Psi}$ and the distribution of $T(\mathbf{s}_m, \mathbf{s}_n, \mathbf{\Psi})$ under $H_1$ is not necessarily uniform. Thus, $c(\alpha)$ enables us to formally control the type-I error rate of our test. The chosen test statistic has some interesting properties:
\begin{enumerate}[1.]
	\item $\mathscr{L}(\mathbf{s}_{m}|\mathbf{s}_n, \mathbf{\Psi})$ decreases as the level of dissimilarity between the trace and control objects increases; hence, $T(\mathbf{s}_m, \mathbf{s}_n, \mathbf{\Psi})$ will tend to 0 as the dissimilarity between trace and control objects increases. Therefore, $T(\mathbf{s}_m, \mathbf{s}_n, \mathbf{\Psi})$ is a strictly positive function and the test defined in (\ref{pvalue}) is a left tail test;
	\item $T(\mathbf{s}_m, \mathbf{s}_n, \mathbf{\Psi})$ only requires $\mathbf{s}_m$ to be random and considers $\mathbf{s}_n$ fixed. This enables the test statistic to be ``anchored'' on the characteristics observed on the control objects sampled from the source considered under $H_1$. In the forensic context, this critical property renders the test specific to the source suspected to have generated the trace fragments. 
\end{enumerate}

\subsection{Accounting for the uncertainty on $\mathbf{\Psi}$ under $H_1$}
\label{psi.uncertainty.H1}
In most situations, $\mathbf{\Psi}$ is not known and must be learned from $\mathbf{s}_n$. \cite{Armstrong:2017} show that an analytical solution to estimate $\mathbf{\Psi}$ from $\mathbf{s}_n$ exists. Instead of replacing $\mathbf{\Psi}$ by a point estimate, $\mathbf{\hat{\Psi}}$, in (\ref{Tfun}), we integrate out the uncertainty associated with the model parameters by considering the posterior distributions of $\theta$, $\sigma^2_a$, and $\sigma^2_e$, given $\mathbf{s}_n$. In this context, we decide to reject $H_1$ if
\begin{eqnarray}
\label{Bayespvalue}
	\int T(\mathbf{s}_m, \mathbf{s}_n, \mathbf{\Psi}) d\pi(\mathbf{\Psi}|\mathbf{s}_n) \ \leq \ c(\alpha).
\end{eqnarray}
The posterior distribution $\pi(\mathbf{\Psi}|\mathbf{s}_n)$ is not a typical Normal-Inverse-Gamma distribution due to the coupling of $\sigma^2_a$ and $\sigma^2_e$ in the covariance matrix of $\mathbf{s}_n$. It is trivial enough to develop a Gibbs sampler to obtain a sample from the distribution. Nevertheless, as we will see in Section \ref{problem1}, it is not necessary. The integral in (\ref{Bayespvalue}) is easily estimated by simulation using Algorithm~\ref{algorithm2}.

\begin{algorithm}[h]
	\setstretch{0.85}
	\caption{Simulation to estimate $\int T(\mathbf{s}_m, \mathbf{s}_n, \mathbf{\Psi}) d\pi(\mathbf{\Psi}|\mathbf{s}_n)$}
	\KwData{A vector of $n+m$ scores}
	\KwResult{Estimate of $\int T(\mathbf{s}_m, \mathbf{s}_n, \mathbf{\Psi}) d\pi(\mathbf{\Psi}|\mathbf{s}_n)$}
	\For{$k \in 1:K$ \normalfont{iterations}}{
		1. Sample $\mathbf{\Psi}^{(k)} := \left\{\sigma^{2(k)}_a, \sigma^{2(k)}_e, \theta^{(k)}\right\}$ from $\pi(\sigma^2_a|\mathbf{s}_n)$, $\pi(\sigma^2_e|\mathbf{s}_n)$, and $\pi(\theta|\mathbf{s}_n, \sigma^2_a, \sigma^2_e)$; \\
		2. Compute the likelihood of the observed scores, $\mathbf{s}_m$, given $\mathbf{\Psi}^{(k)}$, $\mathscr{L}(\mathbf{s}_m|\mathbf{s}_n, \mathbf{\Psi}^{(k)})$; \\
		3. Sample a new vector of scores, $\mathbf{s}_m^{*(k)}$, from $\mathbf{s}^*_m|\mathbf{s}_n, \mathbf{\Psi}^{(k)}$; \\
		4. Compute the conditional likelihood $\mathscr{L}(\mathbf{s}_m^{*(k)} |\mathbf{s}_n, \mathbf{\Psi}^{(k)})$; \\
		5. Determine $L(\mathbf{s}^{*(k)}_m, \mathbf{s}_m,\mathbf{s}_n,\mathbf{\Psi}^{(k)})= \text{I}\left(\mathscr{L}(\mathbf{s}_m |\mathbf{s}_n, \mathbf{\Psi}^{(k)}) \geq \mathscr{L}_k(\mathbf{s}_{m}^{*(k)} |\mathbf{s}_n, \mathbf{\Psi}^{(k)})]\right)$, where $\text{I}(\cdot)$ is the indicator function;	
		}
	Use $\frac{1}{K}\sum_{k=1}^K L(\mathbf{s}^{*(k)}_m, \mathbf{s}_m,\mathbf{s}_n,\mathbf{\Psi}^{(k)})$ to estimate the integral in equation~(\ref{Bayespvalue}). 
	\label{algorithm2}
\end{algorithm}

The output of Algorithm~\ref{algorithm2} converges to $\int T(\mathbf{s}_m, \mathbf{s}_n, \mathbf{\Psi})d\pi(\mathbf{\Psi}|\mathbf{s}_n)$ as $k \rightarrow \infty$, since
\begin{equation}
	\footnotesize
	\begin{aligned}
	\int T(\mathbf{s}_m, \mathbf{s}_n, \mathbf{\Psi})d\pi(\mathbf{\Psi}|\mathbf{s}_n) &= \int \text{Pr}(\mathscr{L}(\mathbf{s}_m^* | \mathbf{s}_n, \mathbf{\Psi}) \leq \mathscr{L}(\mathbf{s}_m | \mathbf{s}_n, \mathbf{\Psi}))d\pi(\mathbf{\Psi}|\mathbf{s}_n)  \\
	&= \int \left[\int \text{I}\left(\mathscr{L}(\mathbf{s}_m^*|\mathbf{s}_n, \mathbf{\Psi}) \leq \mathscr{L}(\mathbf{s}_m|\mathbf{s}_n, \mathbf{\Psi})\right)d\pi(\mathbf{s}_m^*|\mathbf{s}_n, \mathbf{\Psi})\right] d\pi(\mathbf{\Psi}|\mathbf{s}_n)  \\
	&= \int \text{I}\left(\mathscr{L}(\mathbf{s}_m^*|\mathbf{s}_n, \mathbf{\Psi})\leq \mathscr{L}(\mathbf{s}_m|\mathbf{s}_n, \mathbf{\Psi})\right)d\pi(\mathbf{s}_m^*, \mathbf{\Psi}|\mathbf{s}_n)  \\
	&= \underset{k \rightarrow \infty}{lim} \frac{1}{K} \sum_{k=1}^K \text{I}\left(\mathscr{L}(\mathbf{s}_{m_k}^*|\mathbf{s}_n, \mathbf{\Psi}) \leq \mathscr{L}(\mathbf{s}_m|\mathbf{s}_n, \mathbf{\Psi})\right).
	\end{aligned}
\end{equation}
\subsection{Determining $c(\alpha)$}
\label{c.alpha}
In most situations, the distribution of $T(\mathbf{s}_m, \mathbf{s}_n, \mathbf{\Psi})$ may not be uniform since $\mathbf{\Psi}$ is unknown. Therefore, we must determine $c(\alpha)$ empirically. This can be achieved in several ways depending on whether we want to condition $c(\alpha)$ on $\mathbf{s}_n$, or have a decision point that will ensure an average type-I error rate across all possible sources in a population. 

Conditioning $c(\alpha)$ on $\mathbf{s}_n$ implies that the test is specific to the source of the observed control objects. It also presents the advantage that $c(\alpha)$ can be entirely determined by resampling scores using~(\ref{conditional.dist.sm}) and the vector of scores $\mathbf{s}_n$ calculated using the $N$ observed control objects. However, in this situation, $c(\alpha)$ relies heavily on the assumption of normality of the distribution of the scores calculated between objects from the considered source. Furthermore, this strategy assumes that $\mathbf{s}_n$ is a typical sample from its distribution. When $\mathbf{s}_n$ is far from the expectation of its distribution, or when the distribution is not normal, the type-I and II errors of tests conducted using $c(\alpha)$ will vary in unpredictable ways. Alternatively, a source-specific $c(\alpha)$ can be determined by obtaining a very large number of objects from the considered source and using disjoint subsets of these objects to study the empirical distribution of $T(\mathbf{s}_m, \mathbf{s}_n, \mathbf{\Psi})$ under $H_1$. This process has to be repeated for each new test. In most situations, this alternative strategy will be cost-prohibitive. 

The unconditional $c(\alpha)$, obtained using Algorithm~\ref{algorithm4}, has the main advantage that it can be determined for a type of evidence based on a large validation experiment prior to the introduction of the method in casework. By construction, using an unconditional $c(\alpha)$ guarantees that the average type-I error for the considered type of evidence is $\alpha$. However, the type-I error rate cannot be finely controlled for a given specific source. Determining $c(\alpha)$ using this strategy requires samples from a large number of sources. We note that these samples are required to calculate the power of the test, as well as the match probability in the second stage of the approach, and therefore, should be acquired anyway.

\begin{algorithm}[h]
\label{algorithm4}
	\setstretch{0.85}
	\caption{Simulation to determine $c(\alpha)$ across all sources}
	\KwData{A database of $S$ distinct sources}
	\KwResult{Unconditioned $c(\alpha)$ for all sources}
	\For{$k \in 1:K$ \normalfont{iterations}}{
		1. Sample a source $i \in \{1, \dots, S\}$ from the database of sources; \\
		2. Select $N$ control objects from source $i$ in the database; \\
		3. Select $M$ trace objects from source $i$ in the database; \\ 
		4. Compute all pairwise scores, ${\mathbf{s}}_{k}=\binom{{\mathbf{s}}_{m_{k}}}{{\mathbf{s}}_{n_{k}}}$;\\
		5. Use Algorithm~\ref{algorithm2} to approximate $h_{k}=\int T({\mathbf{s}}_{m_{k}}, {\mathbf{s}}_{n_{k}}, \mathbf{\Psi}) d\pi(\mathbf{\Psi}|{\mathbf{s}}_{n_{k}})$; \\
	}
	Define $c(\alpha)$ as the $\alpha-$percentile of the empirical distribution of the $h_{k}$.
\end{algorithm}
As discussed above, there is a fair possibility that the conditional $c(\alpha)$ obtained for a specific source does not correspond to the desired $\alpha$-level of the test. While this possibility also exists with the unconditional $c(\alpha)$, the guarantee that the size of the test is \textit{on average} $\alpha$ for the considered evidence type and the ability to determine $c(\alpha)$ from a large empirical experiment prompts us to recommend the second approach. 

\subsection{Power of the test}
The power of the test introduced in Sections~\ref{alpha_test} and~\ref{psi.uncertainty.H1} cannot be derived analytically given the dimension of the considered objects and the parameter space of the test statistic. However, it can be determined empirically using a reference library of sources that are known to have different characteristics in the input space (e.g., the same collection of sources that is used to determine $c(\alpha)$ in Algorithm~\ref{algorithm4}). Using this library, it is possible to empirically determine the power of the test for fixed numbers of trace and control objects, using Algorithm~\ref{algorithm:POWER}. 

\vspace{3mm}
\begin{algorithm}[H]
	\setstretch{0.85}
	\caption{Simulation to determine the power of the test}
	\KwData{A databases of $S$ distinct sources}
	\KwResult{Power of the test}
	\For{$k \in 1:K$ \normalfont{iterations}}{
		1. Sample a trace source, $i \in \{1, \dots, S\}$, from the database of sources;\\
		2. Sample a control source, $i^* \in \{1, \dots, S\}$, from the database of sources; \\
		3. Sample $M$ trace objects from source $i$ in the database; \\
		4. Sample $N$ control objects from source $i^*$ in the database; \\
		5. Determine the average level of dissimilarity between source $i$ and $i^*$ using the kernel function defined in Section 3.1; \\
		6. Compute all pairwise scores, ${\mathbf{s}}_{k}=\binom{{\mathbf{s}}_{m_{k}}}{{\mathbf{s}}_{n_{k}}}$;\\
		7. Use Algorithm~\ref{algorithm2} to approximate $h_{k}=\int T({\mathbf{s}}_{m_k},{\mathbf{s}}_{n_k}, \mathbf{\Psi}) d\pi(\mathbf{\Psi}|{\mathbf{s}}_{n_k})$;
	}
	Express the $K$ approximations of $h_k$ as a function of $K$ dissimilarities between the trace and control objects.
	\label{algorithm:POWER}
\end{algorithm}
\vspace{3mm}

We stress that the power of our test for a specific $\alpha$-level is not equivalent to the match probability assigned during the second of the two stages of our approach. The power of the test informs on the average probability of erroneously concluding that two sets of objects are indistinguishable as a function of the level of dissimilarity between these two sets. It is determined using sources that are known to have characteristics that are different from each other. The second stage of the approach informs on the case-specific probability that a randomly selected source from a population of potential sources will be a plausible source for the trace objects considered in a case.

\section{Computational considerations}
\label{computation}
Calculating $\frac{1}{K}\sum_{k=1}^K L(\mathbf{s}^{*(k)}_m, \mathbf{s}_m,\mathbf{s}_n,\mathbf{\Psi}^{(k)})$ in Algorithm~\ref{algorithm2} requires posterior samples from $\mathbf{\Psi}|\mathbf{s}_n$, and $\mathbf{s}_m|\mathbf{s}_n,\mathbf{\Psi}^{(k)}$ at each iteration of the algorithm.  We face three challenges when calculating $\frac{1}{K}\sum_{k=1}^K L(\mathbf{s}^{*(k)}_m, \mathbf{s}_m,\mathbf{s}_n,\mathbf{\Psi}^{(k)})$ for large $K$, $M$, or $N$:  
\begin{enumerate}[1.]
	\item Using a Gibbs sampler to obtain a sample from $\pi(\mathbf{\Psi}|\mathbf{s}_n)$ involves a great many number of iterations to obtain a reasonable sample size for $\mathbf{\Psi}$ due to the need to account for the burn-in period and thinning; 
	\item Sampling from $\pi(\theta| \sigma^2_a, \sigma^2_e, \mathbf{s}_n)$ requires calculating the determinant and inverse of $\mathbf{\Sigma}_{n\times n}$ for each new value of $\sigma^2_a$ and $\sigma^2_e$; this may quickly become cumbersome depending on the dimension of $\mathbf{s}_n$ and the number of samples needed;
	\item Similarly, sampling from $\pi(\mathbf{s}_m|\mathbf{s}_n, \mathbf{\Psi}^{(k)})$ requires calculating the determinant and inverse of the conditional covariance matrix $\mathbf{\Sigma}_{m\times m}-\mathbf{\Sigma}_{m \times n}\mathbf{\Sigma}_{n \times n}^{-1}\mathbf{\Sigma}_{n\times m}$ in (\ref{conditional.dist.sm}) for each new sample $\mathbf{\Psi}^{(k)}$; again, this may become a challenge as the dimension of $\mathbf{s}_n$, the dimension of $\mathbf{s}_m$, and the number of samples required increase.
\end{enumerate}
In the following sections, we propose solutions that allow for removing these computational bottlenecks, and enable us to use Algorithm~\ref{algorithm2} with large values for $K$, $M$ and $N$.
\subsection{Posterior sample from $\pi(\mathbf{\Psi}|\mathbf{s}_n)$}
\label{problem1}
Rather than using a Gibbs sampler to obtain posterior samples from $\pi(\mathbf{\Psi}|\mathbf{s}_n)$, we capitalise on the fact that the sums of squares, $SS_a$ and $SS_e$, used in the estimation of $\sigma^2_a$ and $\sigma^2_e$ in \cite{Armstrong:2017} are independent, such that 
\begin{gather*}
	\frac{SS_a}{(N-2)\sigma^2_a+\sigma^2_e} \sim \chi^2_{df=N-1} \hspace{25mm}
	\frac{SS_e}{\sigma^2_e} \sim \chi^2_{df=n-N}.	
\end{gather*}
Defining $\eta_a = (N-2)\sigma^2_a+\sigma^2_e$ and $\eta_e = \sigma^2_e$, we can sample from
\begin{equation}
	\begin{aligned}
		\pi(\eta_a|SS_a, \alpha_a, \beta_a) & \propto \chi^2(SS_a|\eta_a, \alpha_a, \beta_a)\pi(\eta_a|\alpha_a, \beta_a) \\ 
		\pi(\eta_e|SS_e, \alpha_e, \beta_e) & \propto \chi^2(SS_e|\eta_e, \alpha_e, \beta_e)\pi(\eta_e|\alpha_e, \beta_e). \\ 
	\end{aligned}
\end{equation}
Assuming Inverse-Gamma conjugate prior distributions for $\eta_a$ and $\eta_e$, we have that 
\begin{equation}
	\begin{aligned}
		\eta_a | SS_a, \alpha_a, \beta_a  &\sim  IG\left(\alpha_a+\frac{N-1}{2}, \frac{SS_a}{2}+\beta_a\right) \\
		\eta_e | SS_e, \alpha_e, \beta_e &\sim  IG\left(\alpha_e +\frac{n-N}{2}, \frac{SS_e}{2}+\beta_e\right).
	\end{aligned}
\end{equation}
Finally, we can obtain a joint sample of $\sigma^2_a$ and $\sigma^2_e$ from a sample of $\eta_a$ and $\eta_e$ using 
\begin{eqnarray}
	\begin{bmatrix} \sigma^2_a \\ \sigma^2_e \\ \end{bmatrix} = \begin{bmatrix} N-2 & 1 \\ 0 & 1 \end{bmatrix}^{-1} \begin{bmatrix} \eta_a \\ \eta_e \\ \end{bmatrix}.
\end{eqnarray}
Similarly, we can obtain a posterior sample for $\theta$ from a joint sample of $\sigma^2_a$ and $\sigma^2_e$ by assuming a Normal prior with mean and variance parameters $\mu_0$ and $\lambda^2$
\begin{equation}
	\begin{aligned}
		\pi(\theta | \mathbf{s}_n, \sigma^2_a, \sigma^2_e, \mu_0, \lambda^2) &\propto MVN(\mathbf{s}_n|\theta\mathbf{1}_n, \mathbf{\Sigma}_{n\times n}) N(\theta|\mu_0,\lambda^2).
		\label{thetadist}
	\end{aligned}
\end{equation}
The covariance matrix $\mathbf{\Sigma}_{n\times n}$ is a function of $\sigma^2_a$ and $\sigma^2_e$ (Section~\ref{alpha_test}). The parameters $\mu_p$ and $\sigma^2_p$ of the posterior distribution of $\theta$ are equal to
\begin{equation}
	\begin{aligned}
		\mu_p=\frac{\mathbf{1}_n'\mathbf{\Sigma}_{n\times n}^{-1}\mathbf{s}_n\lambda^2+\mu_0}{\left(\mathbf{1}_n'\mathbf{\Sigma}_{n\times n}^{-1}\mathbf{1}_n\right)\lambda^{2}+1}, \hspace{15mm}
		\sigma^2_p = \frac{\lambda^2}{\left(\mathbf{1}_n'\mathbf{\Sigma}_{n\times n}^{-1}\mathbf{1}_n\right)\lambda^2+1}.
		\label{post.params}
	\end{aligned}
\end{equation}
Note that we are not concerned with the choice of the hyperparameters, and that different choices of prior for $\mathbf{\Psi}$ may be considered (e.g., subjective, uninformative, or obtained from the empirical study of a large sample from a population of objects).

This approach allows us to directly generate $i.i.d.$ samples from $\pi(\mathbf{\Psi}| \mathbf{s}_n)$. It does not require a burn-in period or thinning, and therefore does not waste computational resources. However, this approach still requires calculating the determinant and inverse of $\mathbf{\Sigma}_{n \times n}$ for each sample of $\sigma^2_a$ and $\sigma^2_e$ to obtain a new sample of $\theta$. 


\subsection{Determinant and inverse of $\mathbf{\Sigma}_{n\times n}$}
\label{problem2}
We avoid the computational cost of repeatedly inverting $\mathbf{\Sigma}_{n\times n}$ by taking advantage of its spectral decomposition. \cite{Armstrong:2017} show that $\mathbf{\Sigma}_{n\times n}$ has three different eigenvalues
\begin{equation}
	\lambda_{1} =  2\left(N-1\right)\sigma_{a}^{2}+\sigma_{e}, \hspace{10mm} \lambda_{2} =  \left(N-2\right)\sigma_{a}^{2}+\sigma_{e}^{2}, \hspace{10mm} \lambda_{3} =  \sigma_{e}^{2}
\end{equation}
with respective multiplicities 1, $N-1$, and $n-N$. They also show that
\begin{equation}
	\mathbf{\mathbf{\Sigma}_{n\times n}}^{-1} =  \frac{\mathbf{v}_{1}\mathbf{v}_{1}^{t}}{\lambda_{1}}+\sum_{k=2}^{n}\frac{\mathbf{v}_{k}\mathbf{v}_{k}^{t}}{\lambda_{2}}+\sum_{k=n+1}^{N}\frac{\mathbf{v}_{k}\mathbf{v}_{k}^{t}}{\lambda_{3}},
	\label{sigma.inv}
\end{equation}
where $\mathbf{v}_{1}=\frac{\mathbf{1}_{n}}{\sqrt{n}}$ and the $\mathbf{v}_{k}$ are eigenvectors orthogonal to $\mathbf{v}_{1}$, such that
\begin{equation}
	\small
	\begin{aligned}
		\sum_{k=2}^{N}\mathbf{v}_{k}\mathbf{v}_{k}^{t} & = \frac{\left(N-1\right)^{2}}{N-2}\left(\frac{1}{N-1}\mathbf{P}-\frac{1}{n}\mathbf{1}_{n}\mathbf{1}_{N}^{t}\right)\left(\frac{1}{N-1}\mathbf{P}^{t}-\frac{1}{n}\mathbf{1}_{N}\mathbf{1}_{n}^{t}\right) \\
		\sum_{k=N+1}^{n}\mathbf{v}_{k}\mathbf{v}_{k}^{t} & = \mathbf{I}_{n}-\mathbf{v}_{1}\mathbf{v}_{1}^{t} - \sum_{k=2}^{N}\mathbf{v}_{k}\mathbf{v}_{k}^{t}.
		\label{definining.sums}
	\end{aligned}
\end{equation}

Since $N$ is fixed, the general structure of $\mathbf{\Sigma}_{n\times n}$ is fixed. Thus, to obtain $\mathbf{\Sigma}_{n\times n}$ for any new values of $\sigma^2_a$ and $\sigma^2_e$, only the eigenvalues need to be recalculated. This enables us to efficiently obtain the new value for the determinant of $\mathbf{\Sigma}_{n\times n}$ and the inverse of that matrix at each iteration of Algorithm~\ref{algorithm2}. 


\subsection{Resampling from $\mathbf{s}_m|\mathbf{s}_n,\mathbf{\Psi}^{(k)}$}
\label{problem3}
To generate samples of scores from $\mathbf{s}_m|\mathbf{s}_n,\mathbf{\Psi}^{(k)} \sim MVN(\boldsymbol{\mathbf{\mu}}_{\mathbf{s}_m|\mathbf{s}_n,\mathbf{\Psi}^{(k)}}, \mathbf{\Sigma}_{\mathbf{s}_m|\mathbf{s}_n,\mathbf{\Psi}^{(k)}})$ in~(\ref{conditional.dist.sm}), we exploit the properties of the Cholesky decomposition of $\mathbf{\Sigma}_{\mathbf{s}_m|\mathbf{s}_n,\mathbf{\Psi}^{(k)}}=\mathbf{\Sigma}_{m\times m}-\mathbf{\Sigma}_{m \times n}\mathbf{\Sigma}_{n \times n}^{-1}\mathbf{\Sigma}_{n\times m}$. We define $\mathbf{\Sigma}_{\mathbf{s}_m|\mathbf{s}_n,\mathbf{\Psi}^{(k)}}:=\mathbf{L}\mathbf{L}^t$, where $\mathbf{L}$ is a lower triangular matrix. It follows that any vector $\mathbf{s}_{m}^{*(k)} = \boldsymbol{\mathbf{\mu}}_{\mathbf{s}_m|\mathbf{s}_n,\mathbf{\Psi}^{(k)}}+\mathbf{L}\mathbf{z}$, where $\mathbf{z}\in\mathbb{R}^m$ and $z_i \sim N(0,1)$, has mean $\boldsymbol{\boldsymbol{\mathbf{\mu}}}_{\mathbf{s}_m|\mathbf{s}_n,\mathbf{\Psi}^{(k)}}$ and covariance $\mathbf{\Sigma}_{\mathbf{s}_m|\mathbf{s}_n,\mathbf{\Psi}^{(k)}}$, and thus is a sample from  $\pi(\mathbf{s}_m|\mathbf{s}_n,\mathbf{\Psi}^{(k)})$. While $\mathbf{L}$ has to be recalculated for each new sample of $\mathbf{\Psi}^{(k)}$, calculating the Cholesky decomposition of $\mathbf{\Sigma}_{\mathbf{s}_m|\mathbf{s}_n,\mathbf{\Psi}^{(k)}}$ is significantly faster than determining its inverse by other methods.  

\section{Second stage: assigning the \textit{probability of match}}
\label{sec:second.stage}

The focus of the second stage of the approach is to assess the value of finding that the sources of the trace and control objects are indistinguishable from one another (the second stage is not performed when the first stage results in the rejection of the hypothesis of common source at the selected $\alpha$-level). This value is a function of the number of sources, in a population of potential sources, that are also indistinguishable from the source of the trace objects. Thus, the second stage aims at assigning a so-called \textit{probability of match}. Ideally, assigning this probability would require some knowledge of how the characteristics observed on the trace are distributed over the population of potential sources; in turn, this would require defining a likelihood function, which, as mentioned previously, may not exist for most forensic evidence types. 

Instead, for the time being, we propose to follow \cite{Parker:1967}, and repeatedly test whether each source from a sample of sources from a population is indistinguishable from the source of the trace objects using Algorithm~\ref{algorithm2}, keeping $c(\alpha)$ fixed. This process is reflected in Algorithm~\ref{algorithm:RMP}. It allows for using the relative frequency of indistinguishable sources as a proxy for the match probability, as is already typically done for single contributor forensic DNA profiles \citep{Butler:2015}. This process consists in performing multiple dependent $\alpha$-level tests, since the $M$ trace objects are common to all tests. The quality of the relative frequency of indistinguishable sources as an estimate of the match probability depends on how the type-I and type-II errors combine across these different tests. We are currently working on methods to propose a solution to the issue of multiple dependent testing in the context of the proposed model.

\vspace{3mm}
\begin{algorithm}[H]
	\setstretch{0.85}
	\caption{Estimation of the match probability for a set of trace objects}
	\KwData{A set of $M$ trace objects that is decided to be indistinguishable from source $i^*$; a set of sources from the population from which we consider $N$ control objects, $\mathcal{A}:=\{1,2,\dots,S\}\setminus\{i^*\}$; the value of the threshold, $c(\alpha)$, for the considered numbers of trace and control objects}
	\KwResult{An estimate of RMP for a single source}
	\For{$i \in \mathcal{A}$}{	
		Generate a sample of $N$ control objects from source $i$; \\
		Obtain the vector of scores $\mathbf{s}_i=\binom{{\mathbf{s}}_{m_i}}{{\mathbf{s}}_{n_i}}$; \\
		Use Algorithm 1 to estimate $h_i=\int T \left({\mathbf{s}}_{m_i}, {\mathbf{s}}_{n_i}, \boldsymbol{\Psi}\right)d\pi\left(\boldsymbol{\Psi}|{\mathbf{s}}_{n_i}\right)$; \\
	}
	Estimate \textit{RMP}$\ \approx \ \frac{1}{\#\mathcal{A}}\sum_i \text{I}\left(h_{i}>c(\alpha)\right)$, where $\text{I}\left(\cdot\right)$ is the indicator function and $\#\mathcal{A}$ is the cardinality of the set $\mathcal{A}$;
	\label{algorithm:RMP}
\end{algorithm}

\section{Application of the proposed method to paint evidence}
\label{worked.example}
\subsection{Data}
In this section, we apply the proposed approach to Fourier-Transform Infrared spectroscopy (FTIR) spectra of paint chips from cans of common household paint. The paint chips in this example come from 166 different paint cans. For each paint source, we observe seven replicates. Each of the replicates corresponds to a new, distinct observation and is not a repeated measurement on a single paint chip - that is, the seven replicates correspond to seven exchangeable FTIR spectra. Each spectra represents the absorbance of the paint material for a range of wavelengths (from 550 cm$^{-1}$ to 4,000 cm$^{-1}$) and is captured by a 7000-dimensional~vector. 

\begin{figure}[H]
	\centering
	\includegraphics[scale=0.25]{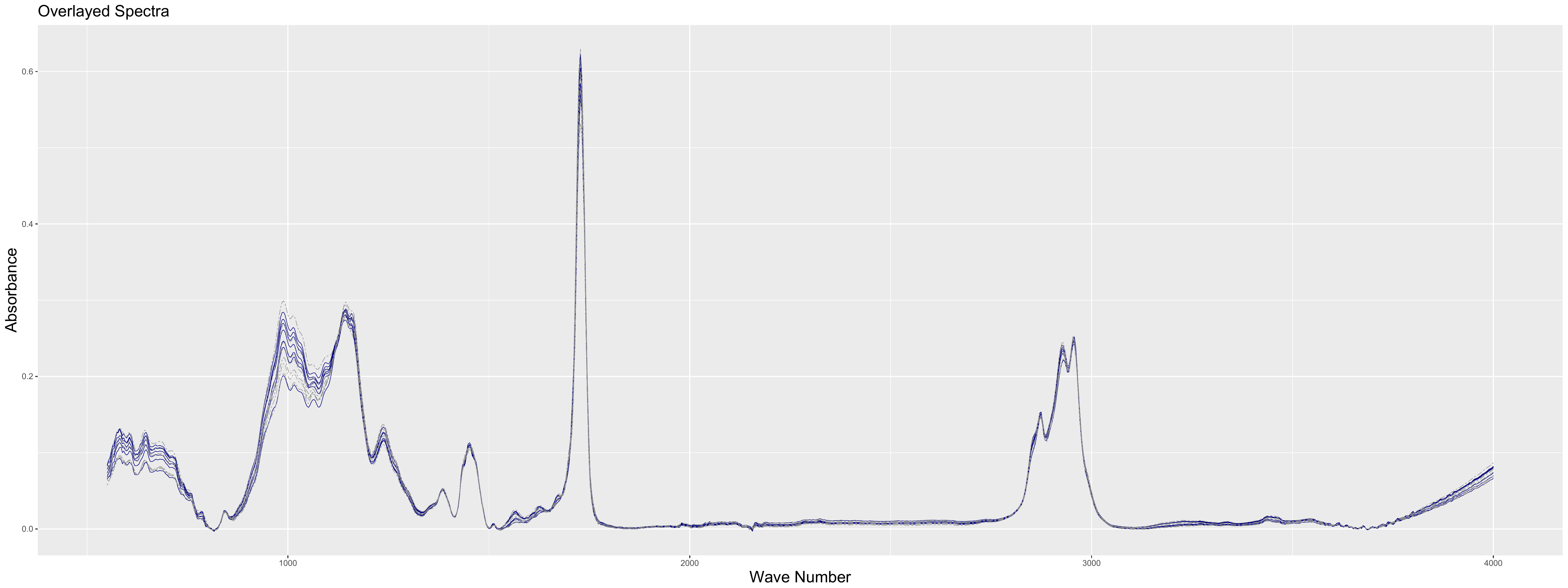}
	\caption{Real Spectra Compared to Pseudo Spectra. Seven observed replicates of spectra (solid dark lines) are overlaid with seven generated replicates of pseudo-spectra (dashed light lines) from source 5.}
	\label{fig:fig1}
\end{figure} 

Since we only observe seven spectra per source, for the purpose of this example, we treat the spectra as functional data and express each one as a linear combination of 300 B-spline bases. We assume that the vectors of basis coefficients are $i.i.d.$ Multivariate Normal, and we use the sample mean and covariance matrix of the coefficients for the seven spectra as point estimates for the parameters of their distribution. This strategy is fit-for-purpose in the context of this example, and enables us to ``resample'' new spectra from a considered source to study the behaviour of our model under different conditions.  Figure~\ref{fig:fig1} shows the reasonableness of this approach. It presents seven observed spectra overlaid with seven simulated spectra from the same can of household paint.

\subsection{Kernel function}
\begin{figure}[H]
	\centering
	\includegraphics[scale=0.25]{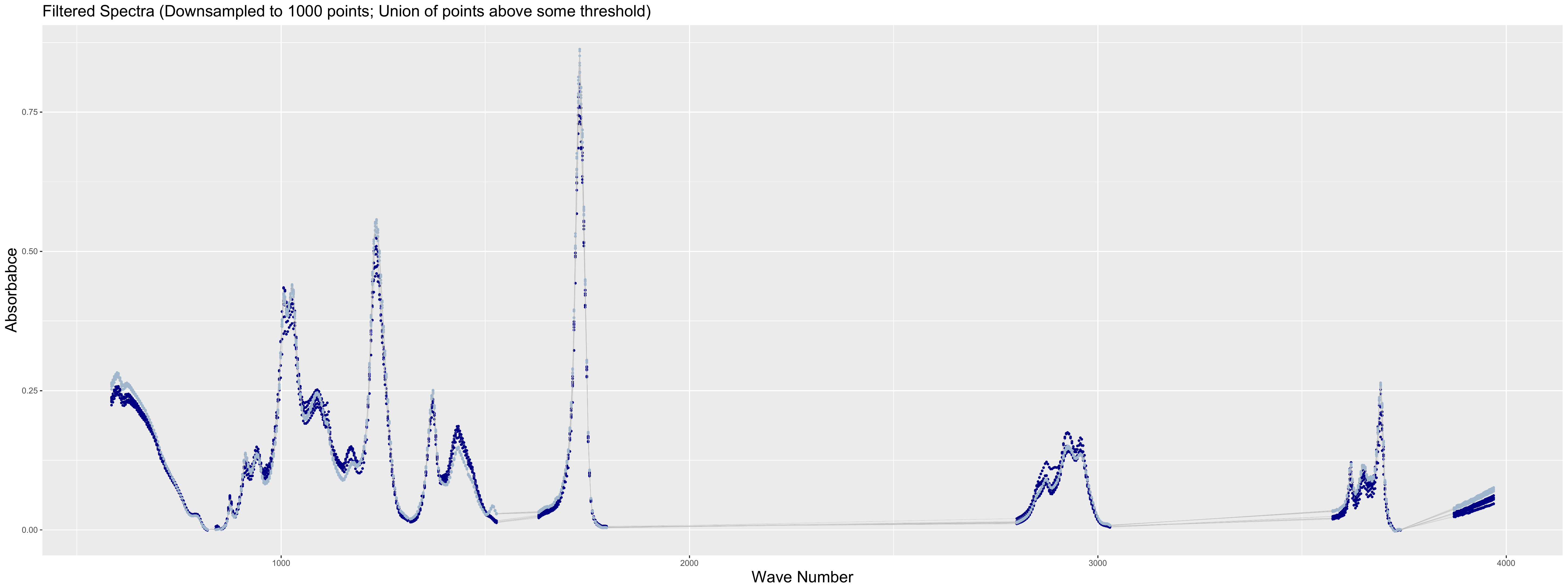}
	\caption{Comparison of Filtered Spectra from Similar Sources. Spectra from source 19 (light blue) overlaid with the most similar source according to the defined kernel function, source 34 (dark blue). Regions in light grey correspond to the areas that are considered to be uninformative in the discrimination process, and so are not considered to calculate the score returned by the kernel function. \label{spec.confused}}
\end{figure}

Our kernel function measures the dissimilarity between two spectra/vectors by considering their cross-correlation (lags -10 to 10) and the Euclidean norm of their difference. As part of the comparison process, the kernel function filters out uninformative areas in a pair of spectra (see Figure~\ref{spec.confused} for an example of the results of the filtering process). Appendix B shows the marginal distributions of score vectors resulting from our kernel function.

\subsection{Determination of $c(\alpha)$}
\begin{figure}[h]
	\centering
	\includegraphics[scale=0.25]{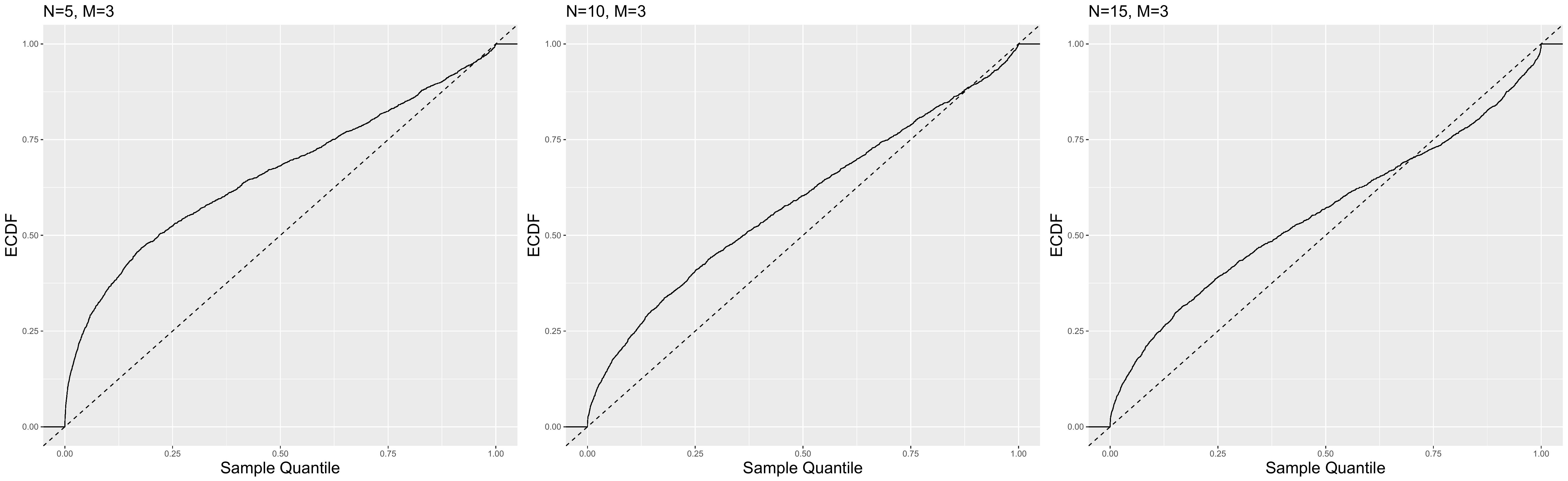}
	\caption{Empirical Distributions of Test Statistics. Empirical cumulative distribution functions (ECDFs) of $\int T(\mathbf{s}_m, \mathbf{s}_n, \mathbf{\Psi}) d\pi(\mathbf{\Psi}|\mathbf{s}_n)$ when $N=5$ and $M=3$ (left), $N=10$ and $M=3$ (middle) and $N=15$ and $M=3$ (right).}
	\label{fig:fig2}
\end{figure}

Figure~\ref{fig:fig2} shows the distribution of the test statistic $\int T(\mathbf{s}_m, \mathbf{s}_n, \mathbf{\Psi})$ $d\pi(\mathbf{\Psi}|\mathbf{s}_n)$ under $H_1$, using the unconditional scenario described in Section~\ref{c.alpha}. 
The three curves correspond to three scenarios where we consider $N=5$, $10$ and $15$ control objects and $M=3$ trace objects. Although each distribution of the test statistic diverges from $Unif(0,1)$, we can use these distribution functions to empirically control the $\alpha$-level of the test (see Table~\ref{calpha.table}). 

\begin{table}[H]
	\caption{Simulation results for obtaining $c(\alpha)$ \label{calpha.table}}
	\begin{center}
	\begin{tabular}{p{1.8cm} p{1.6cm} p{1.45cm} p{1.45cm} p{1.45cm} p{1.45cm} p{1.45cm} p{1.45cm} p{1.45cm}}
	\hline
	$\alpha$ - level & 0.05 & 0.10 & 0.25 & 0.50 & 0.75 & 0.90 & 0.95\\
	\hline
	$c(\alpha)_{N=5}$ & \textbf{0.001895} & 0.006500 & 0.044075 & 0.217650 & 0.621125 & 0.872610 & 0.948015 \\
	$c(\alpha)_{N=10}$ & \textbf{0.006495} & 0.022760 & 0.112025 & 0.361600 & 0.697725 & 0.907480 & 0.966925 \\
	$c(\alpha)_{N=15}$ & \textbf{0.007095} & 0.024390 & 0.116750 & 0.395150 & 0.784650 & 0.943900 & 0.984020 \\
	\hline
	\end{tabular}
	\end{center}
	\footnotesize NOTES: Corresponding $c(\alpha)$'s for various values of $\alpha$ associated with ECDFs of Figure~\ref{fig:fig2} when $N=5$, $N=10$, and $N=15$. Bolded values correspond to those used throughout this example. 
\end{table} 

\subsection{Stage One: Power}

\begin{figure}
	\centering
	\includegraphics[scale=0.25]{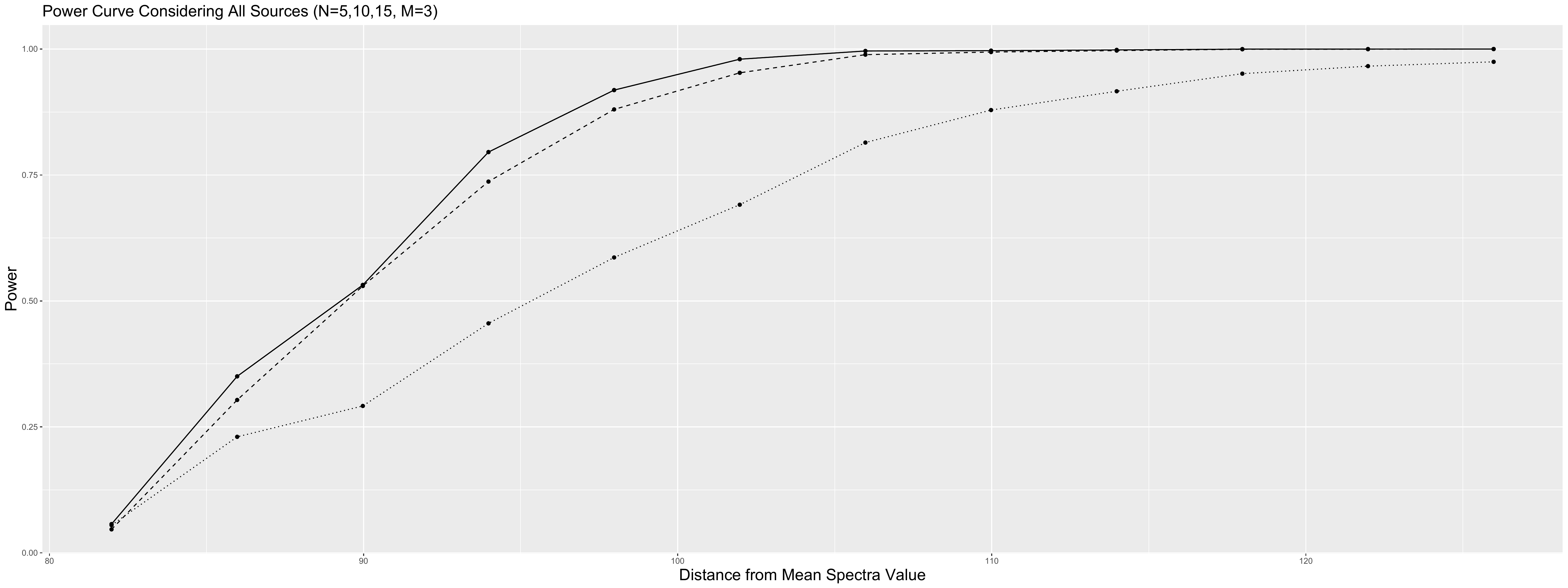}	
	\caption{Power Functions. $P(h\leq c(\alpha)|\kappa(\bar{x}_i, \bar{x}_{i^*}))$ when $N=5$ (dotted line), $N=10$ (dashed line) and $N=15$ (solid line). The x-axis value of the first observation on the left is not to scale. The level of dissimilarity between two very similar spectra is much smaller than represented. We did not observe pairs of sources, in our sample, that had a level of dissimilarity smaller than 85 and we decided to rescale the tail of the power curve for the convenience of the reader.}
	\label{fig:power}
\end{figure}

Figure~\ref{fig:power} presents the power curves associated with the test when $N=5$, $10$, and $15$ as a function of the level of dissimilarity between the average spectra for each source in a pair of sources, $\kappa(\bar{\mathbf{x}}_i, \bar{\mathbf{x}}_{i^*})$, for $i,i^* \in \{1, 2, \dots, S\}$ when $S=166$. Each curve uses the corresponding bolded value of $c(\alpha)$ in Table~\ref{calpha.table} to determine the power as in Algorithm~\ref{algorithm:POWER}. 

Figure~\ref{fig:power} exhibits three typical behaviours of the power function. First, the power of the test approaches one as the distance between two sources increases. Thus, as the characteristics of the trace source become increasingly different from the characteristics of the control source, the test is increasingly able to detect a difference between the sources of the two sets of objects. Second, the power of the test approaches one at a faster rate as the number of control objects increases. Considering a larger set of control objects allows for more precisely assigning the distribution for the within-source comparisons of the control objects, $\mathbf{s}_n$. This consequently improves the ability of the test to differentiate between sets of trace and control objects originating from different sources. Finally, Figure~\ref{fig:power} verifies that the average type-I error for the test is indeed $\alpha$.

\subsection{Stage two: random match probability (RMP)}
\begin{figure}[h]
	\centering
	\includegraphics[scale=0.23]{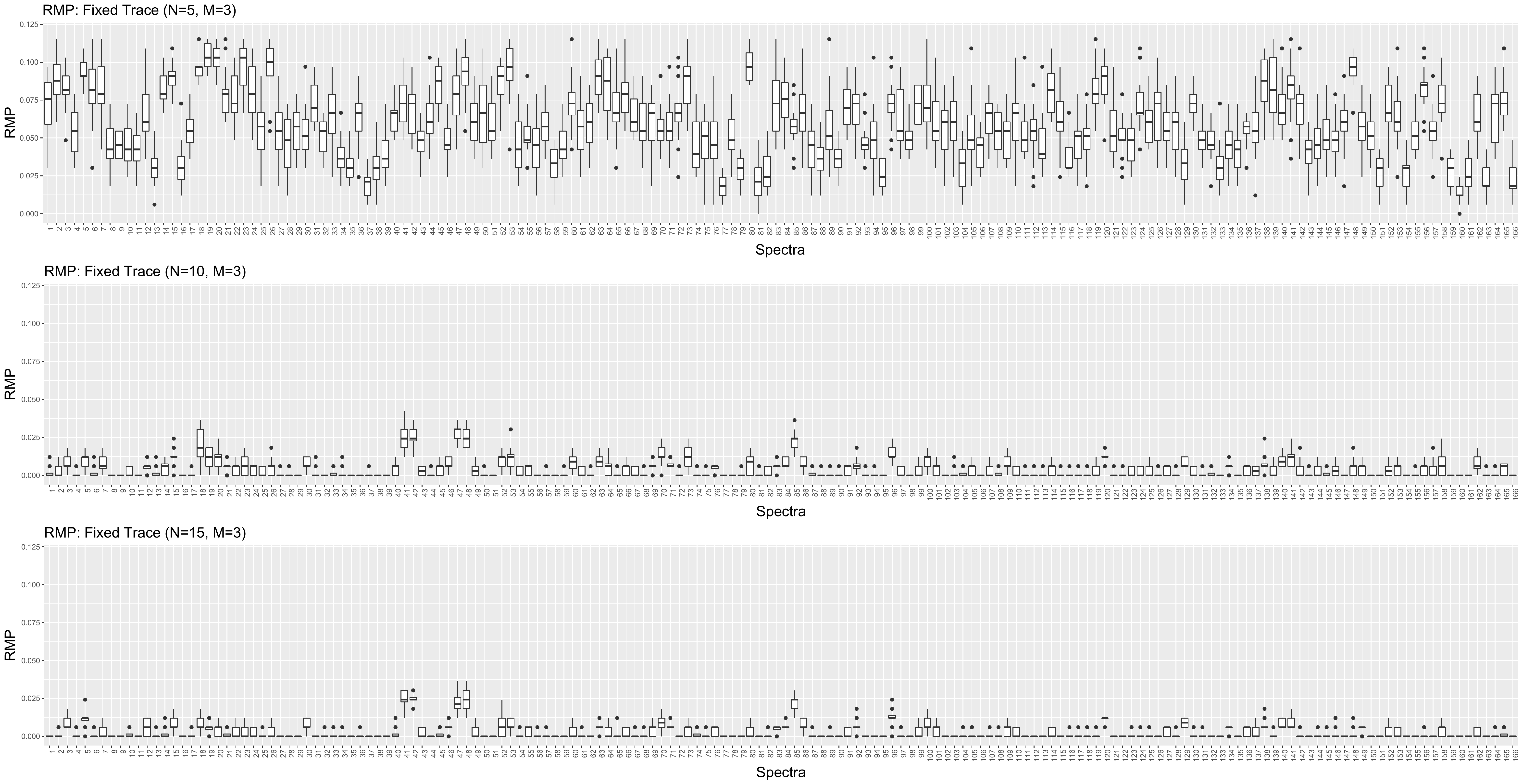}
	\caption{Fixed Trace RMP Distributions. Box plots corresponding to RMP of each source in the population when the M=3 control objects are fixed throughout all simulations when $N=5$ (top), $N=10$ (middle) and $N=15$ (bottom).} \label{img:RMP.fixed}
\end{figure}

The random match probability of any given fixed set of trace objects can be estimated using Algorithm~\ref{algorithm:RMP}. This algorithm considers a fixed set of $M$ trace objects and a randomly sampled set of $N$ control objects for each of the sources representing the population of potential sources. For each source, taken in turn, Algorithm~\ref{algorithm2} is used to test whether the sources of the $M$ and $N$ objects are indistinguishable. The result of each test will be influenced by the random selection of the $N$ control samples for the considered source; hence the RMP estimate for a fixed set of $M$ trace objects may vary with different random sets of control samples from a fixed set of sources. Figure~\ref{img:RMP.fixed} shows the variability between RMP estimates of a unique set of $M=3$ trace objects originating from the source indicated by the x-axis when sets of $N=5$, $10$ and 15 control objects are repeatedly sampled from the other 165 sources.
 Each boxplot represents 20 repetitions of Algorithm~\ref{algorithm:RMP} for the same set of trace objects. 

Several conclusions can be drawn from the data presented in Figure~\ref{img:RMP.fixed}. First, the RMP of trace samples from different sources is not the same (i.e., the locations and spreads of the boxplots vary between different sources). This indicates that some sources of paint appear to have characteristics that are less common in a population of paint than others. Evidence represented by an association between a set of trace objects and control objects from a source displaying such ``rare'' characteristics will carry more weight. Second the median RMP estimate for a unique set of trace objects from a given source is much smaller when sources from the population are represented by $N=10$ and 15 control objects, than when they are represented by $N=5$ control objects. This is a direct result of the observation, in the previous section, that the power of our test increases with the number of control objects. Third, the spread of the RMP estimates for a unique set of trace objects is also much smaller when the sources from the population are represented by $N=10$ and 15 control objects. Greater numbers of observed samples per source imply less uncertainty on the test statistic's parameters, which in turn result in greater precision of the RMP estimates. Finally, increasing the number of control samples from 10 to 15 does not appear to drastically improve the quality of the RMP estimates, despite the significant increase in computational cost. 

\begin{figure}[H]
	\centering
	\includegraphics[scale=0.23]{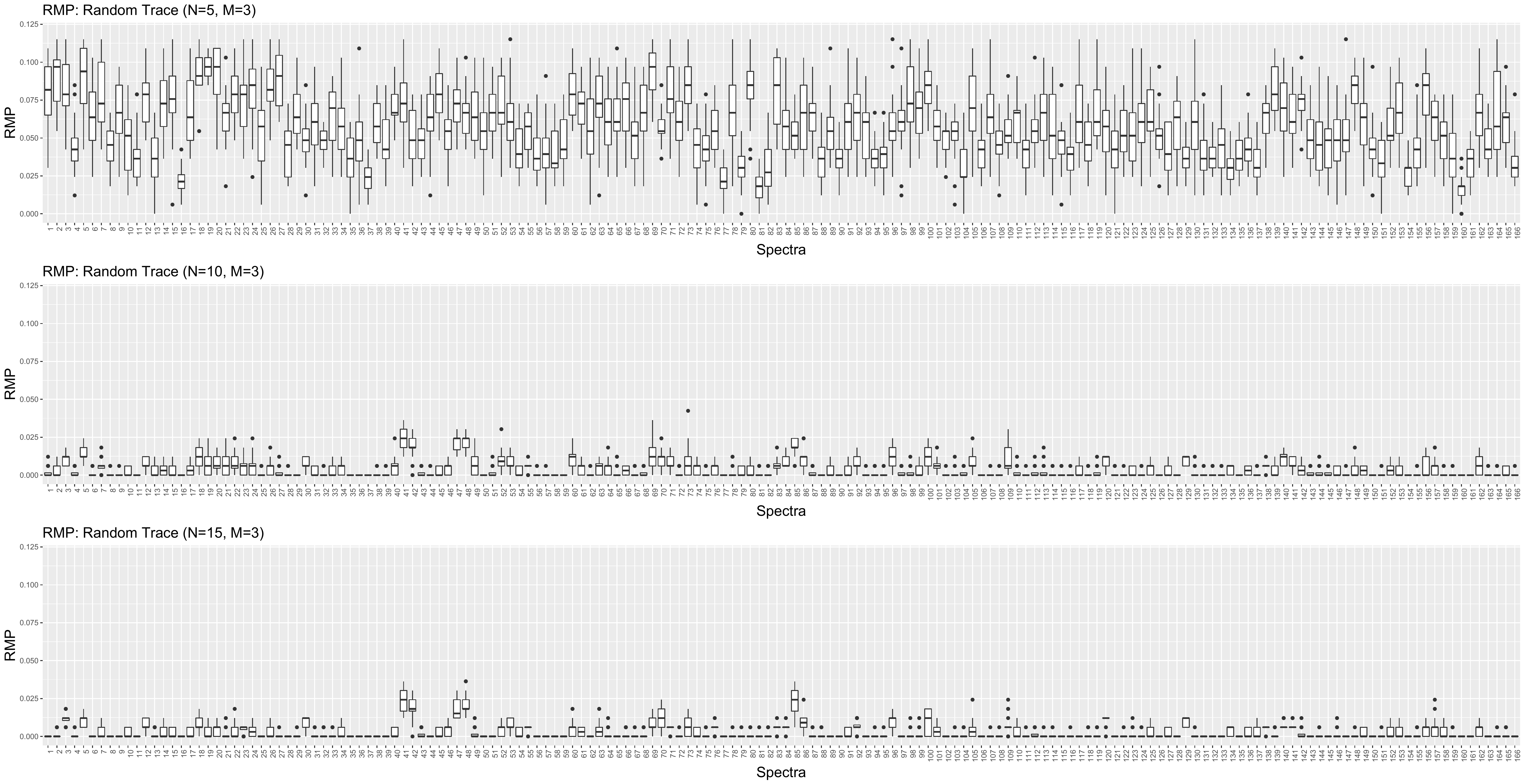}
	\caption{Random Trace RMP Distributions. Box plots corresponding to RMP of each source in the population when the M=3 control objects are randomly sampled for all simulations when $N=5$ (top), $N=10$ (middle) and $N=15$ (bottom).} \label{img:RMP.random}
\end{figure}

Since RMP estimates are conditioned on the observed trace objects, it is possible that the variability in the locations of the RMP estimates presented in Figure~\ref{img:RMP.fixed} is only due to the particular choice of sets of trace objects used in the experiment, and not to the respective rarity of the characteristics of their sources in a population of paint. To test whether the apparent variability in rarity of paint characteristics indicated by the data in Figure~\ref{img:RMP.fixed} is genuine, we repeated the experiment that led to Figure~\ref{img:RMP.fixed}, except that we did not keep the trace objects fixed. The trace objects were randomly sampled for each of the 20 repetitions of Algorithm~\ref{algorithm:RMP}. The results of this experiment are presented in Figure~\ref{img:RMP.random} and confirm the conclusions drawn from the data in Figure~\ref{img:RMP.fixed}.

\section{Benefits and limitations of the two-stage approach}
\label{advantages} 
While the two-stage approach was proposed several decades ago, very little work has been done to formalise and develop it. Nevertheless, this approach has several advantages over the more commonly advocated Bayes factor, and it is not surprising that the value of many evidence types is assessed using some form of (possibly informal) two-stage approach \citep{Aitken:2004}. 

Firstly, the flow of the two-stage approach appears natural to forensic scientists, legal practitioners and lay individuals: (1) the trace and control objects are compared to determine if they could come from the same source; (2) if the two sets of objects are considered \textit{similar}, the implications of this finding is assessed. Each stage focuses on its own specific question. Because the two stages appear so well separated, yet logically connected, they are easy to explain, and even easier to understand by a lay audience, such as a jury~\citep{Neumann:2016}. Scientists can discuss their conclusions for each stage in turn, and how they fit into the overall inference problem. In addition, the issue of error rates naturally occurs in relation to the decision that has to be made at the end of the first stage. The clarity of the two-stage approach to lay individuals has to be put in perspective with the confusion that usually occurs, even among scientists and legal practitioners, when they are asked to use a Bayes factor to update their prior beliefs on the source of trace samples.

Secondly, performing Bayesian model selection using high-dimensional complex data requires using likelihood functions that rely on intractable probability measures in the input space of the data. In addition, data dimension reduction techniques, such as principal component analysis, may engender loss of information that may impact the weight of the evidence in unpredictable ways (e.g., the wrong model may end up being supported at an unknown rate) and may not be applicable to heterogeneous complex data types. On the contrary, it is almost always possible to design a test statistic, study its distribution empirically using a large sample of pairs of objects from the same source, and control the $\alpha$-level of the test, even in the context of high-dimensional complex data. 
   
In this paper, we propose a semi-parametric model that offers several major advantages over a fully empirical approach:
\begin{enumerate}[1.]
	\item Assuming that our decision criterion is an unconditional $c(\alpha)$, for the reasons discussed in the previous section, and that the value of $c(\alpha)$ for the desired $\alpha$-level has been obtained based on a large scale experiment prior to using the test in an operational situation, using the test only requires considering the observations made on the trace and control objects;
	\item The test accounts for the specific characteristics of the source of the control objects;
   	\item The same test statistic can be used for any type or dimension of data. It can be tailored to the data through the use of multiple kernel functions that can be combined ~\citep{Schoelkopf:2001} to maximise the power of the test and ensure that the main assumption of the model, the normality of the score distribution under $H_1$, is satisfied;
   	\item By construction, the test statistic requires only three parameters to be considered, irrespectively of the type and dimension of the raw data. We have shown in the previous sections that this enables us to implement efficient computational strategies to calculate the test statistic and study its distribution under $H_1$ and its power given a suitable sample of sources from a population.
\end{enumerate}


The two-stage approach suffers from several limitations, and, when possible, Bayes factors are preferred to support the inference process. The main objection to the two-stage approach rests in that evidence evaluated using this approach cannot be combined with other pieces of evidence in a logical and coherent manner (see~\cite{Robertson:1995} for a discussion). Another major flaw has been described by \cite{Robertson:1995} as the ``fall-of-the-cliff'' effect (similar to Lindley's paradox~\citep{Lindley:1957}): the decision to reject the hypothesis of a common source during the first stage relies only on whether the value of the test statistic is smaller or larger than a given threshold, and not on the magnitude of the distance between the test statistic and the threshold. Values of the test statistic just beyond the decision threshold will result in a drastically different decision (i.e., exclusion of the considered source) than values just before the threshold (i.e., association of the trace and control objects). In practice, this implies that the source of the control objects is either unequivocally excluded as the source of the trace objects (if $H_1$ is rejected), or that the inference process exclusively favours $H_1$ over $H_2$ (since the match probability of the trace objects in a population of sources will always be lower or equal to 1). By design, the two-stage approach cannot result in a situation where $H_2$ is favoured compared to $H_1$ without $H_1$ being entirely excluded. A further issue with the two-stage approach is related to the power of the test as the quality of the information contained in the trace and control objects decreases. Decreasing quantity and quality of information result in failing to reject $H_1$ at a higher rate. For most applications of statistical hypothesis testing, this would be considered conservative. However, the situation in the forensic context is reversed: failing to reject $H_1$ implies that the suspected source cannot be excluded, and critically, that the inference process will favour the hypothesis that the considered source is in fact the source of the trace versus the hypothesis that the trace material originates from another source in a population of potential sources. In other words, traces with lower quality and quantity of information (i.e., bearing less discriminating features) will be easier to associate to any given suspected source. In the context of the criminal justice system, this behaviour of the test is clearly biased in favour of the prosecution.
\section{Conclusion} 
In this paper, we develop and formalise a two-stage framework for the inference of the source of trace objects in a forensic context. Our approach is particularly useful when the objects are characterised by high-dimensional and complex data, such as chemical spectra, for which likelihood-based inference is not possible. 
	
Although it is not without limitations, the two-stage approach presented in this paper has several major advantages. First, our method provides a framework that enables structured and statistically rigorous inferences in forensic science. The proposed approach may not be as logical and coherent as a fully Bayesian inference framework; however, because its two stages address different and well-defined issues related to the inference process, cognitive research supports that the two-stage approach is a more natural reasoning framework for forensic practitioners and lay individuals alike. 
	
Second, the test statistic and associated likelihood structure proposed in this paper are invariant to the type and dimension of the considered data. The test statistic relies on a kernel function that can be tailored to suit any situation. Thus, the same test statistic can be used in almost any situation where high-dimensional, complex and heterogenous data are considered. In addition, the test's only major assumption can be satisfied through the design of the kernel function, and is naturally satisfied as the dimension of the objects increases.  
	
Much work remains to be done before implementing this methodology in forensic practice. For example, we are developing a model to estimate the match probability in the second stage of the approach and replace the current empirical strategy originally proposed by Parker. Furthermore, large reference collections of different types of evidence (e.g., paint, fibers, glass) need to be gathered. 

The application of our method to FTIR data of paint shows two important results: FTIR spectra of paint contain highly specific information that enable discrimination of paint samples from different sources; and the characteristics of some paint sources are rarer than others and will carry more probative value. Our results do not only show that paint evidence is very probative in general, but they also show that our approach works well with the number of samples typically encountered in casework. Our approach can easily be implemented to determine the probative value of paint evidence in any given case, hence addressing the recurrent criticisms related to the lack of quantitative support for forensic conclusions. Finally, our approach can easily be extended to other evidence types, such as transferred automotive paint in road accidents, transferred glass fragments during burglaries, assaults or shootings and transferred fibres from items of clothing during assaults.  

\bibliographystyle{chicago}

\clearpage
\setcounter{equation}{0}
\setcounter{table}{0}
\setcounter{figure}{0}
\renewcommand{\theequation}{A.\arabic{equation}}
\renewcommand{\thetable}{A.\arabic{table}}
\renewcommand{\thefigure}{A.\arabic{figure}}

\section*{Appendix A: Summary of the main results from Armstrong et al. (2017)}
\label{armstrong}

Given two vectors of measurements $\mathbf{x}_i$ and $\mathbf{x}_j$ representing the observations made on two objects, $i,j$, sampled from a common source, a kernel function, $\kappa$, is used to measure their level of similarity and report it as a score, $s_{i,j}$. The score is represented by a linear random effects model 
\begin{eqnarray}
	s_{i,j}=\kappa(\mathbf{x}_i,\mathbf{x}_j)=\theta+a_i+a_j+\epsilon_{i,j},
	\label{single.score.model}
\end{eqnarray}
where $\theta$ is the expected value of the score between any two objects from the same considered source; $a_i$, $a_j$ are random effects representing the contributions of the $i$th and $j$th objects, and $\epsilon_{ij}$ is a lack of fit term, such that $a_i$ and $a_j\stackrel{iid}\sim N(0,\sigma^2_a)$, and $\epsilon_{ij}\sim N(0,\sigma^2_e)$. 

We note that the kernel function at the core of the model, $\kappa$, can be designed to accommodate virtually any type of data. It needs satisfy only two requirements: it must be a symmetric function, that is $\kappa(\mathbf{x}_i,\mathbf{x}_j)=\kappa(\mathbf{x}_j,\mathbf{x}_i)$; and it must ensure that the marginal distribution of $s_{ij}$ is Normal to satisfy the assumption made on the score model in~(\ref{single.score.model}). The assumption of normality is the main assumption made by \cite{Armstrong:2017} when developing their model; it is reasonable for high-dimensional objects and can be satisfied through careful design of the kernel function \citep{ArmstrongPhD}.

The vector of all possible pairwise comparisons between $N$ reference objects can be represented by a vector, $\mathbf{s}_n$ of $n=\binom{N}{2}$ objects, given by $\mathbf{s}_n=(s_{1,2}, s_{1,3}, \dots, s_{n-1,n})^t$. The multivariate extension of the model in (\ref{single.score.model}) to $\mathbf{s}_n$ is given by
\begin{eqnarray}
	\mathbf{s}_n=\theta\mathbf{1}_n+\mathbf{P}\mathbf{a}+\mathbf{\epsilon},
	\label{multivariate.extension}
\end{eqnarray}
where $\mathbf{1}_n$ is a one vector of length $n$, $\mathbf{P}$ is an $n\times N$ design matrix (where each row represents an $i,j$ combination, consisting of ones in the $i^{th}$ and $j^{th}$ columns and zeros elsewhere), $\mathbf{a}$ is the vector of random effects for the considered objects, and $\mathbf{\epsilon}$ is the vector of $\epsilon_{ij}$ corresponding to each pair of objects. By construction, 
\begin{eqnarray}
	\mathbf{s}_n\sim MVN(\theta\mathbf{1}_n,\mathbf{\Sigma}_{n\times n}) \text{, where} \hspace{3mm} \mathbf{\Sigma}_{n\times n}=\mathbf{PP}^t\sigma^2_a+\mathbf{I}_n\sigma^2_e. 
	\label{sn.dist}
\end{eqnarray}
\cite{Armstrong:2017} show that $\mathbf{\Sigma}_{n\times n}$ has three different eigenvalues
\begin{eqnarray}
		\lambda_{1} =  2\left(N-1\right)\sigma_{a}^{2}+\sigma_{e}; \hspace{15mm}
		\lambda_{2} =  \left(N-2\right)\sigma_{a}^{2}+\sigma_{e}^{2} \hspace{15mm}
		\lambda_{3} =  \sigma_{e}^{2}
		\label{eigenvals}
\end{eqnarray}
with multiplicity 1, $N-1$, and $N-n$ respectively. \cite{Armstrong:2017} also show that
\begin{equation}
\begin{aligned}
	|\mathbf{\Sigma}_{n\times n}| & =  \left(2\left(n-1\right)\sigma_{a}^{2}+\sigma_{e}^{2}\right)\left(\left(n-2\right)\sigma_{a}^{2}+\sigma_{e}^{2}\right)^{\left(n-1\right)}\left(\sigma_{e}^{2}\right)^{N-n}\\
	\mathbf{\mathbf{\Sigma}_{n\times n}}^{-1} & =  \frac{\mathbf{v}_{1}\mathbf{v}_{1}^{t}}{\lambda_{1}}+\sum_{k=2}^{n}\frac{\mathbf{v}_{k}\mathbf{v}_{k}^{t}}{\lambda_{2}}+\sum_{k=n+1}^{N}\frac{\mathbf{v}_{k}\mathbf{v}_{k}^{t}}{\lambda_{3}}
	\end{aligned}
	\label{sigma.inv}
\end{equation}
where $\mathbf{v}_{1}=\frac{\mathbf{1}_{n}}{\sqrt{n}}$ and $\mathbf{v}_{k}$ are eigenvectors orthogonal to $\mathbf{v}_{1}$. Importantly, \cite{Armstrong:2017} note that 

\begin{equation}
	\begin{aligned}
		\sum_{k=2}^{N}\mathbf{v}_{k}\mathbf{v}_{k}^{t} & = \frac{\left(N-1\right)^{2}}{N-2}\left(\frac{1}{N-1}\mathbf{P}-\frac{1}{n}\mathbf{1}_{n}\mathbf{1}_{N}^{t}\right)\left(\frac{1}{N-1}\mathbf{P}^{t}-\frac{1}{n}\mathbf{1}_{N}\mathbf{1}_{n}^{t}\right) \\
		\sum_{k=N+1}^{n}\mathbf{v}_{k}\mathbf{v}_{k}^{t} & = \mathbf{I}_{n}-\mathbf{v}_{1}\mathbf{v}_{1}^{t} - \sum_{k=2}^{N}\mathbf{v}_{k}\mathbf{v}_{k}^{t}
		\label{definining.sums}
	\end{aligned}
\end{equation}

Using these results, \cite{Armstrong:2017} show that the likelihood function, $\mathscr{L}\left(\theta,\sigma_{a}^{2},\sigma_{e}^{2}|\mathbf{s}_n\right)$, can be rewritten as an independent sum of squares
\begin{eqnarray}
	-2\ \mathscr{L}\left(\theta,\sigma_{a}^{2},\sigma_{e}^{2}|\mathbf{s}_n\right) & = & \log\left(\lambda_{1}\right)+\left(N-1\right)\log\left(\lambda_{2}\right)+\left(n-N\right)\log\left(\lambda_{3}\right)+n\log\left(2\pi\right)\nonumber \\
 	& + & \frac{n\left(\bar{s}_n-\theta\right)^{2}}{\lambda_{1}}+\frac{SS_{a}}{\lambda_{2}}+\frac{SS_{e}}{\lambda_{3}}.
 	\label{log.lik}
\end{eqnarray}
where $\bar{s}_n$ is the average of the elements in $\mathbf{s}_n$, and
\begin{equation}
	\begin{aligned}
		SS_a & = \frac{\left(N-1\right)^{2}}{N-2}\sum_{i=1}^{N}\left(\bar{s}_n^{\left(i\right)}-\bar{s}_n\right)^{2}, \\
		SS_e & = \mathbf{s}_{n}^{t}\left(\mathbf{I}_{n}-\mathbf{v}_{1}\mathbf{v}_{1}^{t}\right)\mathbf{s}_{n} - SS_a,
		\label{sums.squares}
	\end{aligned}
\end{equation}
where $\bar{s}_n^{(i)}$ is the average of the elements in $\mathbf{s}_n$ involving object $i$.

Finally, \cite{Armstrong:2017} show that closed-form solution estimates for the parameters of the model exist and can be derived from Table~\ref{ANOVA.table} to obtain
\begin{eqnarray}
		\hat{\theta} = \bar{s} \hspace{20mm}
		\hat{\sigma}^2_a = \frac{MS_a-MS_e}{N-2} \hspace{20mm}
		\hat{\sigma}^2_e = MS_e.
		\label{point.estimates}
\end{eqnarray}

\begin{table}[h]
	\caption{ANOVA table for the model $\mathbf{s}_n\sim MVN(\theta\mathbf{1}_n, \mathbf{\Sigma}_{n\times n})$ \label{ANOVA.table}}
	\begin{center}
	\begin{tabular}{p{1.6cm} p{1.6cm} p{1.6cm} p{1.6cm} p{3cm}}
	\hline\hline
	\em{Source}&\em{df}&\em{SS}&\em{MS}&\em{E(MS)}\\   [1ex]
	\hline\hline
	A & $N-1$ & $SS_a$ & $\frac{SS_a}{(N-1)}$ & $(N-2)\sigma^2_a + \sigma^2_e$\\
	Error & $n-N$ & $SS_e$ & $\frac{SS_e}{(n-N)}$ & $\sigma^2_e$\\
	Total & $n-1$ & $SS_t$ & $\frac{SS_t}{(n-1)}$ & \\
	\hline
	\end{tabular}
	\end{center}
\end{table}
At this point, we simply note that we presented the results obtained by \cite{Armstrong:2017} for a set of $N$ objects known to come from a single source and that it is trivial to scale these results for a vector containing the pairwise scores resulting from the cross-comparisons of $N+M$ objects, if they are assumed to originate from the same source. 

\section*{Appendix B: Multivariate normality of scores used in the application of our method}
\label{normalityassumption}

\begin{figure}[h]
	\centering
	\includegraphics[scale=0.40]{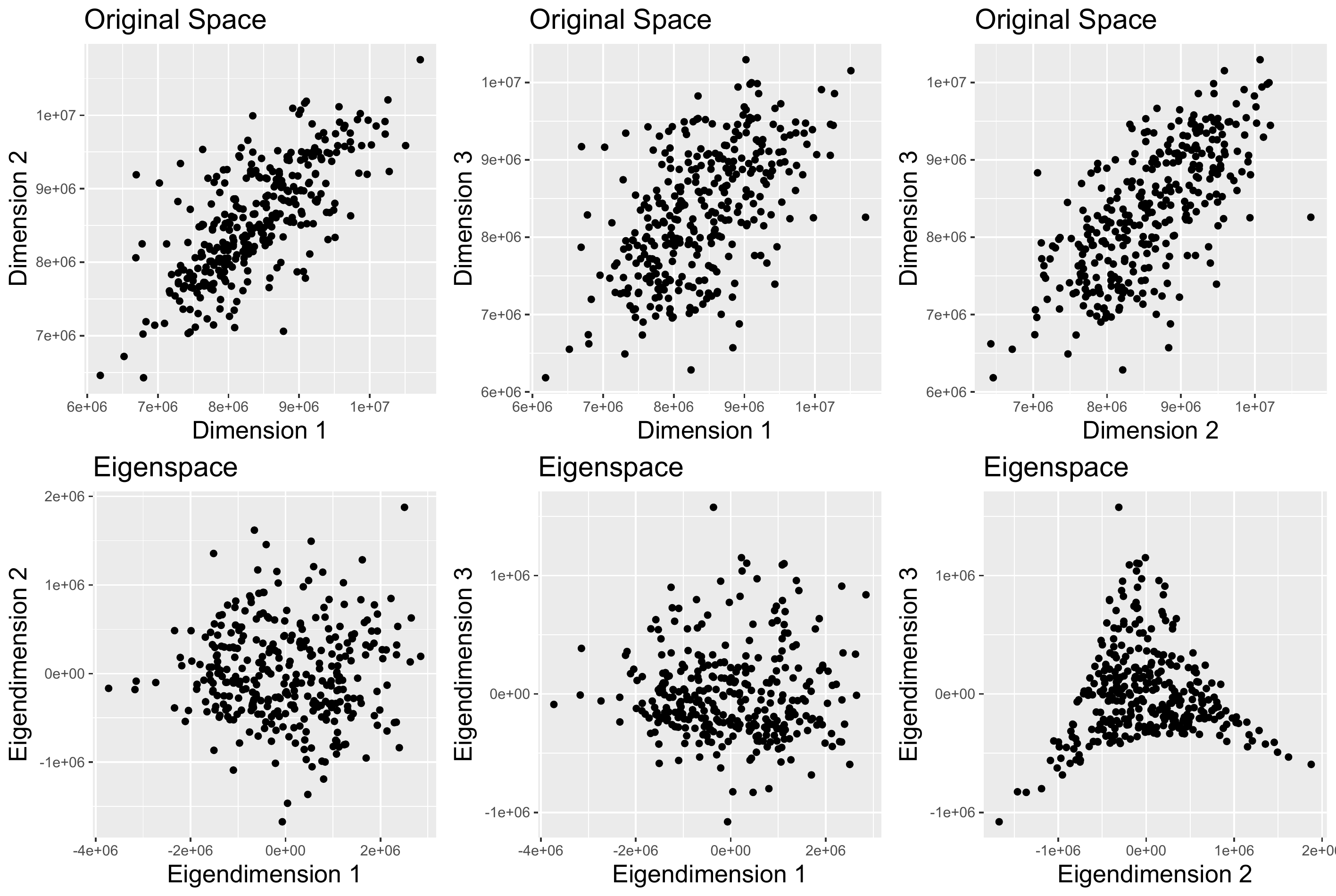}
	\caption{Original and Projected Distributions of Scores. Top Row: 3-dimensional vectors of scores obtained from 332 triplets of objects originating from the same source in the original space. Bottom Row: Projection of 3-dimensional vectors of scores obtained from 332 triplets of objects originating from the same source in the space defined by the spectral decomposition of their covariance matrix. \label{eigenscores}}
\end{figure}

To examine whether the scores considered in this paper approximately satisfy the assumption of multivariate normality of the score model proposed by~\cite{Armstrong:2017}, we consider 332 triplicates of spectra originating from the same source. These 332 triplicates consist in two disjoint triplicates of spectra from each of the 166 paint sources described in Section~\ref{worked.example}. The top row of Figure~\ref{eigenscores} portrays the marginal distributions of the scores in their original space. By expressing the original 3-dimensional vectors of scores as a function of the space defined by the eigenvectors of their sample covariance matrix, we can observe the marginal distributions of the score vectors along orthogonal axes, and better determine if the marginal distributions follow a Normal distribution. Figure~\ref{eigenscores} (bottom row) shows that, although the data is approximately spherical in the first two dimensions of the eigenspace, there is a rather significant departure from normality when eigendimensions 2 and 3 are plotted against each other. Given the results in Section~\ref{worked.example}, we purport that this deviation from multivariate normality does not affect the ability of the model to correctly classify and differentiate spectra, and thus testifies to the robustness of the model: despite the lack of normality, the model is still able to correctly associate and differentiate spectra originating from the same and different sources, respectively.  

%
%

\end{document}